\documentclass[useAMS,usenatbib]{mn2e}

\usepackage[psamsfonts]{amssymb}
\usepackage[dvips]{graphicx}
\usepackage{amsmath,alltt}
\usepackage{multirow}
\usepackage{rotating}
\usepackage{lscape}

\title[Blue Straggler Formation in Globular Clusters]{An Analytic Model for
  Blue Straggler Formation in Globular Clusters}
\author[Nathan Leigh, Alison Sills and Christian Knigge]{Nathan
  Leigh$^{1}$,
  Alison Sills$^{1}$, Christian Knigge$^{2}$\thanks{E-mail:
    leighn@mcmaster.ca (NL);
    asills@mcmaster.ca (AS); christian@astro.soton.ac.uk (CK)} \\
$^{1}$Department of Physics and Astronomy, McMaster University,
1280 Main St. W., Hamilton, ON, L8S 4M1, Canada \\
$^{2}$School of Physics and Astronomy, University of Southampton,
Highfield, Southampton, SO17 1BJ, United Kingdom}
\begin{document}

\pagerange{\pageref{firstpage}--\pageref{lastpage}} \pubyear{2010}

\maketitle

\label{firstpage}

\begin{abstract}
We present an analytic model for blue straggler formation in 
globular clusters.  We assume that blue stragglers are formed
only through stellar collisions and binary star evolution, and
compare our predictions to observed blue straggler numbers taken from the
catalogue of \citet{leigh11a}.  We can summarize our key results as
follows:  (1) Binary star evolution consistently dominates blue
straggler 
production in all our best-fitting models.  (2) In order to account
for the observed sub-linear dependence of blue 
straggler numbers on the core masses \citep{knigge09}, the core binary
fraction must be inversely proportional to the total cluster
luminosity and should always exceed at least a few percent.  (3) In at
least some clusters, blue straggler formation must be enhanced by
dynamical encounters 
(either via direct collisions or by stimulating mass-transfer to occur
by altering the distribution of binary orbital parameters)
relative to what is expected by assuming a simple population of
binaries evolving in isolation.  (4) The agreement between the
predictions of our model and the observations can be improved 
by including blue stragglers that form outside the core but
later migrate in due to dynamical friction.  (5) Longer blue
straggler lifetimes are preferred in models that include blue
stragglers formed outside the core since this increases the fraction
that will have sufficient time to migrate in via dynamical friction.  
\end{abstract}


\section{Introduction} \label{intro}

Commonly found in both open and globular clusters (GCs), blue
stragglers (BSs) appear as an
extension of the main-sequence (MS) in cluster colour-magnitude
diagrams (CMDs), occuyping the region that is just brighter and bluer
than the main-sequence turn-off (MSTO) 
\citep{sandage53}.  BSs are
thought to be produced via the addition of hydrogen to 
low-mass MS stars \citep[e.g.][]{sills01, lombardi02}.  This can
occur via multiple channels, most of which involve the mergers of
low-mass MS stars since a significant amount of mass is typically
required to reproduce the observed locations of BSs in CMDs
\citep[e.g.][]{sills99}.  Stars in close binaries can merge if enough
orbital angular momentum is lost, which can be mediated by dynamical
interactions with other stars, magnetized stellar winds, tidal
dissipation or even an outer triple companion
\citep[e.g.][]{leonard92, li06, perets09, dervisoglu10}.
Alternatively, MS stars can collide directly, although this is
also thought to usually be mediated by multiple star systems
\citep[e.g.][]{leonard89, leonard95, fregeau04, leigh11b}.  First proposed by
\citet{mccrea64}, BSs have also been hypothesized to form by
mass-transfer from an evolving primary onto a normal MS companion
via Roche lobe overflow.

Despite numerous formation
mechanisms having been proposed, a satisfactory explanation to account
for the presence of BSs in star clusters eludes us still.  Whatever the
dominant BS 
formation mechanism(s) operating in dense clusters, it is now
thought to somehow involve multiple star systems.  This was shown
to be the case in even the dense cores of GCs \citep{leigh07, leigh08,
  knigge09} where
collisions between single stars are thought to occur frequently
\citep{leonard89}.  In \citet{knigge09}, we showed that the numbers of
BSs in the cores
of a large sample of GCs correlate with the core masses.  We
argued that our results are consistent with what is expected if BSs
are descended from binary stars since this would imply a dependence of
the form $N_{BS} \sim f_bM_{core}$, where $N_{BS}$ is the number of
BSs in the core, $f_b$ is the binary fraction in the core and
$M_{core}$ is the total stellar mass contained within the core.
\citet{mathieu09} also showed 
that at least $76\%$ of the BSs in the old open cluster NGC 188 have
binary companions.  Although the nature of these companions remains
unknown, it is clear that binaries played a role in the
formation of these BSs.  

Blue stragglers are typically concentrated in the dense cores of
globular clusters where the high 
stellar densities should result in a higher rate of stellar encounters 
\citep[e.g.][]{leonard89}.  Whether or not this fact is
directly related to BS formation remains unclear, since mass segregation 
also acts to migrate BSs (or their progenitors) into the core 
\citep[e.g.][]{saviane98, guhathakurta98}.  Additionally, numerous BSs
have been observed in more sparsely populated open clusters
\citep[e.g.][]{andrievsky00} and the fields of GCs where
collisions are much less likely to occur and mass-transfer within
binary systems is thought to be a more likely formation scenario
\citep[e.g.][]{mapelli04}.  

Several studies have provided evidence
that BSs show a bimodal spatial distribution in some GCs
\citep{ferraro97, ferraro99, lanzoni07}.  In these clusters, the BS
numbers are the highest in the central cluster 
regions and decrease with increasing distance from the cluster centre
until a second rise occurs in the cluster outskirts.  
This drop in BS numbers at intermediate cluster radii is often
referred to as the ``zone of avoidance''.  Some authors have suggested
that it is the result of two separate formation mechanisms
occurring in the inner and outer regions of the cluster, with
mass-transfer in primordial binaries dominating in the latter and
stellar collisions dominating in the former
\citep{ferraro04, mapelli06}.  Conversely, mass segregation could also
give rise to a ``zone of avoidance'' for BSs if the time-scale for dynamical
friction exceeds the average BS lifetime in only the outskirts of GCs that
exhibit this radial trend \citep[e.g.][]{leigh11a}. 

Dynamical interactions occur frequently enough in dense clusters that
they are expected to be at least partly responsible for the observed
properties of BSs \citep[e.g.][]{stryker93, leigh11b}.  It follows
that the current properties of BS populations
should reflect the dynamical histories of their host clusters.  As a
result, BSs could provide an indirect means of probing
the physical processes that drive star cluster evolution
\citep[e.g.][]{heggie03, hurley05, leigh11b}.

In this paper, our
goal is to constrain the dominant BS formation mechanism(s) operating
in the dense cores of GCs by analyzing the principal processes
thought to influence their production.  To this end, we
use an analytic treatment to obtain predictions for the number of BSs
expected to be found within one core radius of the cluster centre at
the current cluster age.  Predicted numbers for the core are calculated
for a range of free parameters, and then compared to the
observed numbers in order to find the
best-fitting model parameters.  In this way, we are able to quantify
the degree to which each of the considered formation mechanisms
should contribute to the total predicted numbers in order to
best reproduce the observations.  

In Section~\ref{data}, we describe the BS catalogue used for comparison
to our model predictions.  In 
Section~\ref{method}, we present our analytic model for BS 
formation as well as the statistical technique we have developed to
compare its predictions to the observations.  
These predictions are then compared to the observations in
Section~\ref{results} for a range of model parameters.  In
Section~\ref{discussion}, we discuss the implications of our results
for BS formation, as well as the role played by the cluster dynamics
in shaping the current properties of BS populations.

\section{The Data} \label{data}

The data used in this study was taken from \citet{leigh11a}.  In that 
paper, we presented a catalogue for blue straggler, red giant branch
(RGB), horizontal branch (HB) and main-sequence turn-off stars
obtained from the colour-magnitude diagrams of 35 Milky Way GCs
taken 
from the ACS Survey for Globular Clusters \citep{sarajedini07}.  The
ACS Survey provides unpecedented deep photometry in the F606W ($\sim$
V) and F814W ($\sim$ I) filters
that extends reliably from the HB all the way down to about 7
magnitudes below the MSTO.  The clusters in our sample span a range of
total masses (by nearly 3 orders of magnitude) and central
concentrations \citep{harris96}.  We have confirmed that 
the photometry is nearly complete in the BS region of the CMD for
every cluster in our sample.  This was 
done using the results of artificial star tests taken from
\citet{anderson08}.

Each cluster was centred in the ACS field, which 
extends out to several core radii from the cluster
centre in most of the clusters in our sample.  Only the core
populations provided in \citet{leigh11a} are used in this paper.
We have taken estimates for the core radii and central luminosity
densities for the clusters in our sample from
\citet{harris96}, whereas central velocity 
dispersions were taken from \citet{webbink85}.  Estimates for the
total stellar mass contained within the core were obtained from
single-mass King models, as described in \citet{leigh11a}.  All of the
clusters in our sample were chosen to be non-post-core collapse, and
have surface brightness profiles that provide good fits to our King
models. 

\section{Method} \label{method}

In this section, we present our model and outline our
assumptions.  We also present the statistical technique used to
compare the 
observed number counts to our model predictions in order to identify
the best-fitting model parameters. 

\subsection{Model} \label{model}

Consider a GC core that is home to N$_{BS,0}$ BSs at some time t $=$
t$_0$.  At a specified time in the future, the number of BSs in the
core can be approximated by: 
\begin{equation}
\label{eqn:number-bss}
N_{BS} = N_{BS,0} + N_{coll} + N_{bin} + N_{in} - N_{out} - N_{ev},
\end{equation}
where N$_{coll}$ is the number of BSs formed from collisions during
single-single (1+1), single-binary (1+2) and binary-binary (2+2)
encounters, N$_{bin}$ is the number formed from 
binary evolution (either partial mass-transfer between the binary
components or their complete coalescence), N$_{in}$ is the number 
of BSs that migrate into the core due to dynamical friction, N$_{out}$
is the number that migrate out of the core via kicks experienced during
dynamical encounters, and N$_{ev}$ is the number of BSs that have
evolved away from being brighter and bluer than the MSTO in the
cluster CMD due to stellar evolution.  

We adopt an average stellar mass of $m = 0.65 M_{\odot}$ and an
average BS mass of $m_{BS} = 2m = 1.3 M_{\odot}$.  The mass of a BS
can provide a rough guide to its lifetime, although a range of
lifetimes are still possible for any given mass.  
For instance, \citet{sandquist97} showed that a
1.3 M$_{\odot}$ blue straggler will have a lifetime of around $0.78$
Gyrs in unmixed models, or $1.57$ Gyrs in 
completely mixed models.  Combined with the results of \citet{sills01},
\citet{lombardi02} and \citet{glebbeek08}, we expect a lifetime in the
range 1-5 Gyrs for a 1.3 M$_{\odot}$ BS.  As a first approximation, we
choose a likely value of $\tau_{BS} = 1.5$ Gyrs for the average BS
lifetime \citep[e.g.][]{sills01}.  The effects had on our
results by changing our assumption for the average BS lifetime will be
explored in Section~\ref{results} and discussed in
Section~\ref{discussion}. 

We consider only the last $\tau_{BS}$ years.  This is because we are
comparing our model predictions to current
observations of BS populations, so that we are only concerned with
those BSs formed within the last few Gyrs.  Any BSs formed before this
would have evolved away from being brighter and bluer than the MSTO by
the current cluster age.  Consequently, we set N$_{BS,0}$ =
N$_{ev}$ in Equation~\ref{eqn:number-bss}.  We further assume that
all central cluster parameters have not changed in the last
$\tau_{BS}$ years, including the central velocity dispersion, the
central luminosity density, the core radius and the core binary
fraction.  It follows that the rate of BS formation is constant for
the time-scale of interest.  This time-scale is
comparable to the half-mass relaxation time but much longer than the
central relaxation time for the majority of the clusters in our sample
\citep{harris96}.  This suggests that core
parameters such as the central density and the core radius will
typically change in a time $\tau_{BS}$ since the time-scale on
which these parameters vary is the central relaxation
time \citep{heggie03}.  Therefore, our assumption of constant rates
and cluster parameters is not strictly correct, however it provides a
suitable starting point for our model.  We will discuss the
implications of our assumption of time-independent cluster properties
and rates in Section~\ref{discussion}.

In the following sections, we discuss each of the remaining terms in
Equation~\ref{eqn:number-bss}.

\subsubsection{Stellar Collisions} \label{collisions}

We can approximate the number of BSs formed in the last $\tau_{BS}$
years from collisions during dynamical encounters as:  
\begin{equation}
\label{eqn:number-coll}
N_{coll} = f_{1+1}N_{1+1} + f_{1+2}N_{1+2} + f_{2+2}N_{2+2},
\end{equation}
where N$_{1+1}$, N$_{1+2}$ and N$_{2+2}$ are the number of
single-single, single-binary and binary-binary encounters,
respectively.  The terms f$_{1+1}$, f$_{1+2}$ and f$_{2+2}$ are the
fraction of 1+1, 1+2 and 2+2 encounters, respectively, that will
produce a BS in the last $\tau_{BS}$ years.  We treat these three
variables as free parameters since we do not know what fraction of
collision products will produce BSs (i.e. stars with an appropriate
combination of colour and brightness to end up in the BS region of the
CMD), nor do we know what fraction of 1+2 and 2+2 encounters will
result in a stellar collision.  Numerical scattering experiments have
been performed to study the outcomes of 1+2 and 2+2 encounters
\citep[e.g.][]{hut83, mcmillan86, fregeau04}, however a large fraction
of the relevant parameter space has yet to be explored.

In terms of the core radius $r_c$ (in parsecs), the central number
density $n_0$ (in pc$^{-3}$), the root-mean-square velocity $v_{m}$
(in km s$^{-1}$), the average stellar mass $m$ (in M$_{\odot}$) and
the average stellar radius $R$ (in R$_{\odot}$), the mean time-scale
between single-single collisions in the core of a GC is
\citep{leonard89}:
\begin{equation}
\begin{gathered}
\label{eqn:coll1+1}
\tau_{1+1} = 1.1 \times 10^{10}(1-f_b)^{-2} \Big(\frac{1 pc}{r_c}
\Big)^3 \Big(\frac{10^3 pc^{-3}}{n_0} \Big)^2 \\
 \Big(\frac{v_{m}}{5
  km/s} \Big) \Big(\frac{0.5 M_{\odot}}{m} \Big) \Big(\frac{0.5
  R_{\odot}}{R} \Big)\mbox{ years}
\end{gathered}
\end{equation}
The additional factor (1-f$_b$)$^{-2}$ comes from the fact that we
are only considering interactions between single stars and the
central number density of single stars is given by (1-f$_b$)n$_0$,
where f$_b$ is the binary fraction in the core (i.e. the fraction of
objects that are binaries).  For our chosen mass, we assume a
corresponding average stellar radius using the relation M/M$_{\odot}$
= R/R$_{\odot}$ \citep{iben91}.
The number of 1+1 collisions expected to have occurred in the last
$\tau_{BS}$ years is then approximated by:
\begin{equation}
\label{eqn:N-1+1}
N_{1+1} = \frac{\tau_{BS}}{\tau_{1+1}}.
\end{equation}

The rate of collisions
between single stars and binaries, as well as between two
binary pairs, can be roughly approximated in the same way
as for single-single encounters \citep{leonard89, sigurdsson93,
  bacon96, fregeau04}.  We adopt the time-scales derived in
\citet{leigh11b} for the average times between 1+2 and 2+2 encounters.
These are:
\begin{equation}
\begin{gathered}
\label{eqn:coll1+2}
\tau_{1+2} = 3.4 \times 10^7f_b^{-1}(1-f_b)^{-1}\Big(\frac{1 pc}{r_c}
\Big)^3 \Big(\frac{10^3 pc^{-3}}{n_0} \Big)^2 \\
 \Big(\frac{v_{m}}{5
  km/s} \Big) \Big(\frac{0.5 M_{\odot}}{m} \Big) \Big(\frac{1 AU}{a}
\Big)\mbox{ years}
\end{gathered}
\end{equation}
and
\begin{equation}
\begin{gathered}
\label{eqn:coll2+2}
\tau_{2+2} = 1.3 \times 10^7f_b^{-2}\Big(\frac{1 pc}{r_c}
\Big)^3 \Big(\frac{10^3 pc^{-3}}{n_0} \Big)^2 \\ 
\Big(\frac{v_{m}}{5
  km/s} \Big) \Big(\frac{0.5 M_{\odot}}{m} \Big) \Big(\frac{1 AU}{a}
\Big)\mbox{ years},
\end{gathered}
\end{equation}
where $a$ is the average binary semi-major axis in the core in AU and
we have assumed that the average binary mass is equal to twice the
average single star mass.  
The numbers of 1+2 and 2+2 encounters expected to have occurred in the 
last $\tau_{BS}$ years are given by, respectively:
\begin{equation}
\label{eqn:N-1+2}
N_{1+2} = \frac{\tau_{BS}}{\tau_{1+2}}
\end{equation}
and
\begin{equation}
\label{eqn:N-2+2}
N_{2+2} = \frac{\tau_{BS}}{\tau_{2+2}}.
\end{equation}

The outcomes of 1+2 and 2+2 encounters will ultimately contribute to
the evolution of the binary fraction in the core.  How and with what
frequency binary 
hardening/softening as well as capture, exchange and ionization
interactions affect the binary fraction in the dense cores of GCs is
currently a subject of debate \citep[e.g.][]{ivanova05,
  hurley07}.  
Observations are also lacking for binary fractions in the dense cores of
GCs, however rough constraints suggest that they range from a few to a
few tens of a percent \citep[e.g.][]{rubenstein97, cool02, sollima08,
  davis08}.  The 
situation is even worse for the distribution of binary orbital
parameters observed in dense stellar environments.  Our best
constraints come from radial velocity surveys of moderately dense open
clusters \citep{latham05, geller09}, however whether or not the
properties of the binary populations in these clusters should differ
significantly from those in the much denser cores of GCs is unclear.
As an initial assumption, we assume a time-independent core binary
fraction of 10\% for all clusters, and an average semi-major axis
of 2 AU.  This semi-major axis corresponds roughly to the hard-soft
boundary for most of the clusters in our sample, defined by setting
the average binary orbital energy equal to the kinetic energy of an
average star in the cluster \citep[e.g.][]{heggie03}.  We 
treat both the core binary 
fraction and the average binary semi-major axis as free parameters,
and explore a range of possibilities using the available observations
as a guide for realistic values.  We will return to these assumptions 
in Section~\ref{discussion}.

\subsubsection{Binary Star Evolution} \label{mergers}

Although we do not know the rate of BS formation from binary star
evolution, we expect a general relation of the form
N$_{bin}$ = $\tau_{BS}$/$\tau_{mt}$ for the number of BSs produced
from binary mergers in the last $\tau_{BS}$ years, where $\tau_{mt}$
is the average time between BS formation events due to binary star
evolution.  We can express the number of BSs formed from binary star
evolution in the last $\tau_{BS}$ years as:
\begin{equation}
\label{eqn:N-bin}
N_{bin} = f_{mt}f_bN_{core},
\end{equation}
where  N$_{core}$ is the total number of
objects (i.e. single and binary stars) in the core and f$_{mt}$ is
the fraction of binary stars that evolved internally to form
BSs within the last $\tau_{BS}$ years.  We treat f$_{mt}$ as a free
parameter since it is likely to depend on the mass-ratio, period and
eccentricity distributions characteristic of the binary populations of
evolved GC cores, for which data is scarce at best.  

\subsubsection{Migration Into and Out of the Core}
\label{segregation}

Due to the relatively small sizes of the BS populations
considered, the migration of BSs into or out of the core is an
important consideration when calculating the predicted numbers.  In
other words, we are dealing with relatively small
number statistics and every blue straggler counts.  In order to
approximate the number of stars in the core as a function of time,
two competing effects need to be taken into account:  (1) mass
stratification/segregation (or, equivalently, dynamical friction) and
(2) kicks experienced during dynamical interactions.  
These effects are accounted for with the variables N$_{in}$ and
N$_{out}$ in Equation~\ref{eqn:number-bss}, respectively.  

Blue stragglers are among the most massive stars in clusters
\citep[e.g.][]{shara97, vandenberg01, mathieu09}, so they should
typically be heavily mass segregated relative to other stellar
populations \citep[e.g.][]{spitzer69, shara95, king95}.  The
time-scale for this process to occur 
can be approximated using the dynamical friction time-scale \citep{binney87}:
\begin{equation}
\label{eqn:t-dyn}
\tau_{dyn} = \frac{3}{4ln{\Lambda}G^2(2{\pi})^{1/2}}\frac{\sigma(r)^3}{m_{BS}\rho(r)},
\end{equation}  
where $\sigma(r)$ and $\rho(r)$ are, respectively, the velocity
dispersion and stellar mass density at the given distance from the
cluster centre $r$.  Both $\sigma(r)$ and $\rho(r)$ are found from
single-mass King models \citep{sigurdsson93}, which are fit to each
cluster using the 
concentration parameters provided in \citet{mclaughlin05}.  
The Coulomb logarithm is denoted by $\Lambda$, and we adopt
a value of ln$\Lambda \sim 10$ throughout this paper
\citep[e.g.][]{spitzer87, heggie03}.  If $\tau_{dyn} >
\tau_{BS}$ at a given distance from the 
cluster centre, then any BSs formed at this radius in the last
$\tau_{BS}$ years will not have had sufficient time to migrate into
the core by the current 
cluster age.  The maximum radius r$_{max}$ at which BSs can
have formed in the 
last $\tau_{BS}$ years and still have time to migrate into the core
via dynamical friction is given by setting $\tau_{dyn} = \tau_{BS}$.  
Therefore, N$_{in}$ depends only on the number of BSs formed in the last
$\tau_{BS}$ years at a distance from the cluster centre smaller than
r$_{max}$.  

In order to estimate the contribution to N$_{BS}$ in
Equation~\ref{eqn:number-bss} from BSs formed outside the core, we
calculate the number of BSs formed in radial shells between the
cluster centre and r$_{max}$.  Each shell is taken to be one core
radius thick, and we calculate the contribution from 
each formation mechanism in every shell.  This is done by assuming a
constant (average) density and velocity dispersion in each shell.
Specifically, we estimated the density and velocity 
dispersion at the half-way point in each shell using our single-mass King models,
and used these to set average values.  The number of BSs expected to
have migrated into the core within the last $\tau_{BS}$ years can be
written:
\begin{equation}
\begin{gathered}
\label{eqn:N-in}
N_{in} = \sum_{i=2}^N \Big( f_{1+1}N_{(1+1),i} + f_{1+2}N_{(1+2),i} +
f_{2+2}N_{(2+2),i} \\ 
+ f_{mt}N_{(bin),i} \Big) \times \Big(1 -
\frac{\tau_{(dyn),i}}{\tau_{BS}}\Big),
\end{gathered}
\end{equation}
where $i=1$ refers to the core, $i=2$ refers to the shell immediately 
outside the core, etc. and N is taken to be the integer nearest to
r$_{max}$/r$_c$.  We let the terms with N$_{(1+1),i}$, N$_{(1+2),i}$,
N$_{(2+2),i}$ and 
N$_{(bin),i}$ represent the number of BSs formed in shell $i$ from
single-single collisions, single-binary collisions, binary-binary
collisions and binary star evolution, respectively.  The time-scale
for dynamical friction in shell $i$ is denoted by $\tau_{(dyn),i}$, and
the factor (1 - $\tau_{(dyn),i}$/$\tau_{BS}$) is included to account for
the fact that we are assuming a constant formation rate for BSs, so
that not every BS formed in shells outside the core will have
sufficient time to fall in by the current cluster age.

It is typically the least massive stars
that are ejected from 1+2 and 2+2 interactions as single stars
\citep[e.g.][]{sigurdsson93}.  Combined with conservation of momentum,
this suggests that BSs are the least likely to be ejected from
dynamical encounters with velocities higher than the central velocity
dispersion due to their large masses.  This has been confirmed by
several studies of numerical scattering experiments 
\citep[e.g.][]{hut83, fregeau04}.  Based on this, we 
expect that N$_{out}$ should be 
very small and so, as a first approximation, we take N$_{out} = 0$.
However, we also explore the effects of a 
non-zero N$_{out}$ by assigning a kick velocity to all BSs at birth.
If dynamical interactions play a role in BS formation, we 
might naturally expect BSs to be imparted a recoil velocity at birth
(or shortly before) due to 
momentum conservation.  We will return to this assumption in
Section~\ref{results}. 

\subsection{Statistical Comparison with Observations} \label{statistics}

Our model contains 4 free parameters, which correspond to the fraction of
outcomes that produce a blue straggler for each formation mechanism
(1+1 collisions, 1+2 collisions, 2+2 collisions, and binary star
evolution).  These are the $f$ values
described in the previous section: $f_{1+1}, f_{1+2}, f_{2+2},
f_{mt}$.  We assume that these values are constant throughout each
cluster, and are also constant between clusters.  By fitting the
predictions of our model to the observational data, we can determine
best-fit values for each of these $f$ parameters, and therefore make
predictions about which blue straggler formation processes are
more important. 

In order to determine the best values for these $f$ parameters, we 
need an appropriate statistical test.  For this, we follow the approach
of \citet{verbunt08}.  The number of BSs seen in the core of
a globular cluster can be described by Poisson statistics.  In
particular, the probability of observing $N$ sources when $\mu$ are
expected is:
\begin{equation}
\label{eqn:stats}
P(N,\mu) = \frac{\mu^N}{N!}e^{-\mu}
\end{equation}
We can calculate a probability for each cluster, and then calculate an 
overall probability $P^{\prime}$ for the model by multiplying the
individual $P$ values.  We can then vary the $f$ values to maximize
this value. 

In practice, these $P$ values are typically of order ten percent per
cluster, and with 35 clusters, the value of $P^{\prime}$ quickly
becomes extremely small.  Therefore we chose to work with a modified
version of this value:  the deviance of our model to the saturated
model.  A saturated model is one in which the observed number of
sources is exactly equal to the expected number in each cluster.  In
other words, this is the best that we can possibly do.  However,
because of the nature of Poisson statistics, the probability $P$ of
such a model (calculated by setting $N=\mu$ in
Equation~\ref{eqn:stats}) is not equal to 1, but in fact has some
smaller value.  For the numbers of blue stragglers in our clusters,
the $P$ values for the saturated model run from 0.044 to 0.149, and
the value of $P^{\prime}$ is $2.08 \times 10^{-41}$.

The deviance of any model from the saturated model is given by
\begin{equation}
\label{eqn:deviance}
D = 2.0(\ln(P^{\prime}_{saturated}) - \ln(P^{\prime}_{model}))
\end{equation}
The model which minimizes this quantity will be our best-fit
model.  Ideally, the scaled deviance ($D/(N_{data}-N_{parameters}$)
should be equal to 1 for a best fit.  Given the simplicity of our
model, we expect that our values will not provide this kind of
agreement, and we simply look for the model which provides the minimum
of the scaled deviance.

\section{Results} \label{results}

In this section, we present the results of comparing our model
predictions to the observations.  After presenting the results for 
a constant core binary fraction for all clusters, we explore the
implications of adopting a core binary fraction that depends on the
cluster luminosity, as reported in \citet{sollima07} and
\citet{milone08}. 

\subsection{Initial Assumptions} \label{initial}

\begin{figure}
\begin{center}
\includegraphics[width=\columnwidth]{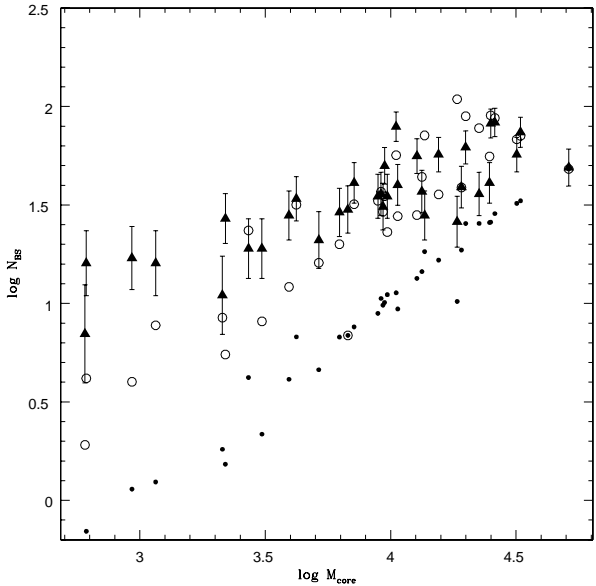}
\end{center}
\caption[The predicted number of BSs plotted versus the total stellar
mass in the core for the best-fitting model parameters found for our
initial choice of assumptions]{
The predicted number of BSs plotted versus the total stellar
mass in the core for the best-fitting model parameters found for our
initial choice of assumptions.  The filled triangles correspond to
  the observed numbers, the 
  open circles to the number of BSs predicted to have formed within
  r$_{max}$, and the small filled circles to the number predicted to
  have formed in only the core.  The best-fitting model parameters
  used to calculate the predicted numbers are f$_{1+1} = 0$, f$_{1+2} = 0$,
  f$_{2+2} = 7.3 \times 10^{-3}$, and f$_{mt} = 1.7 \times 10^{-3}$.  
  Estimates for the total stellar masses in the core were obtained
  from single-mass King models, as described in \citet{leigh11a}.
  Error bars have been indicated for the observed numbers using 
  Poisson statistics.
\label{fig:model_best}}
\end{figure}

The predictions of our model for our initial choice of assumptions are
shown in Figure~\ref{fig:model_best}.  These numbers are plotted
against the total stellar mass in the core along with the 
number of BSs observed in the core (filled triangles).  We plot both the
total number of BSs predicted to have formed within $r_{max}$ in the
last $\tau_{BS}$ years (N$_{BS}$ in Equation~\ref{eqn:number-bss};
open circles), as well as the total number formed only in the core
(N$_{coll}$ + N$_{bin}$; small filled circles).  Upon comparing N$_{BS}$ to the
observed number of BSs in the core, 
the best-fitting model parameters predict that most BSs are formed
from binary star evolution, with a small contribution from 2+2
collisions being needed in order to obtain the best possible match to
the observations.  The ideal contribution from 2+2 collisions
constitutes at 
most a few percent of the predicted total for most of the clusters in
our sample.  Even for our best-fitting model
parameters, our initial choice of assumptions predicts too few BSs in
clusters with small core masses. 

\subsection{Binary Fraction} \label{results-binary}

\begin{figure}
\begin{center}
\includegraphics[width=\columnwidth]{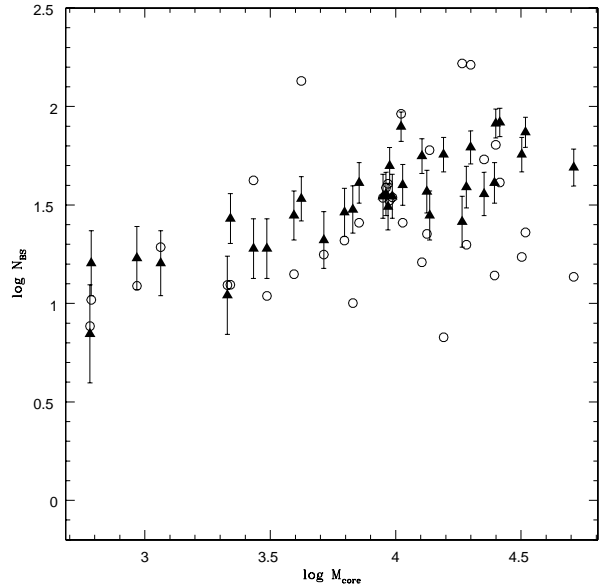}
\end{center}
\caption[The predicted number of BSs plotted versus the total stellar
mass in the core for the best-fitting model parameters found using the
binary fractions of \citet{sollima07} with f$_b^{min} = 0.01$]{The
  number of BSs predicted in the cluster core using the binary
  fractions of \citet{sollima07} with f$_b^{min} = 0.01$ plotted
  against the total stellar mass contained within the core.  The
  symbols used to indicate the observed and predicted numbers are the
  same as in Figure~\ref{fig:model_best}.  The predicted numbers correspond to
  the best-fitting model parameters, which are f$_{1+1} = 0.41$,
  f$_{1+2} = 0$, f$_{2+2} = 0$, and f$_{mt} = 1.5 \times 10^{-3}$.
\label{fig:model_rmax_sollima_best}}
\end{figure}

We tried changing our assumption of a constant f$_b$ for all
clusters to one for which the core binary fraction varies with the
total cluster magnitude.  First, we adopted a dependence of the
form:
\begin{equation}
\label{eqn:sollima07} 
f_b = 0.13M_V + 1.07,
\end{equation}
where M$_V$ is the total cluster V magnitude.  This 
relation comes from fitting a line of best-fit to the observations of 
\citet{sollima07}, who studied the binary fractions in a sample
of 13 low-density GCs (we calculated an average of
columns 3 and 4 in their Table 3 and used these binary fractions to
obtain Equation~\ref{eqn:sollima07}).  In order to prevent negative
binary fractions, 
we impose a minimum binary fraction of f$_b^{min} = 0.01$.  In other
words, we set f$_b =$ f$_b^{min}$ if Equation~\ref{eqn:sollima07}
gives a binary fraction less than f$_b^{min}$.  As before, we adopt an
average semi-major axis of 2 AU.  The results of this comparison are
presented in Figure~\ref{fig:model_rmax_sollima_best}.  As in
Figure~\ref{fig:model_best}, both the numbers of 
observed (filled triangles) and predicted (open circles) BSs in the
core are plotted 
versus the total stellar mass in the core.  Once again, the predicted
numbers include all BSs formed within r$_{max}$ in the last
$\tau_{BS}$ years.  The best-fitting model parameters for this
comparison suggest that both single-single collisions and binary star
evolution are significant contributors to BS formation.  Single-single
collisions contribute up to several tens of a percent of the predicted
total in several clusters.  We obtain a 
deviance in this case that is significantly larger than that obtained
for the best-fitting model parameters assuming a constant f$_b$ of
$10\%$ for all clusters.  Equation~\ref{eqn:sollima07} gives higher
binary fractions in low-mass
clusters relative to our initial assumption of a constant f$_b$.  This
increases the number of BSs formed from binary star evolution and
improves the agreement between our model predictions and the
observations in low-mass clusters.  This is consistent with the
results of \citet{sollima08}.  However, adopting
Equation~\ref{eqn:sollima07} for f$_b$ also causes our model to
under-predict the number of BSs in several high-mass clusters.

\begin{figure}
\begin{center}
\includegraphics[width=\columnwidth]{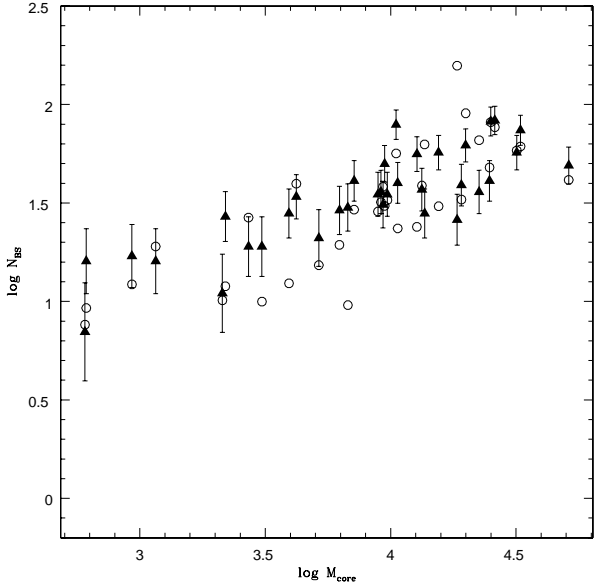}
\end{center}
\caption[The predicted number of BSs plotted versus the total stellar
mass in the core for the best-fitting model parameters found using the binary
fractions of \citet{sollima07} with f$_b^{min} = 0.1$]{The number of BSs
  predicted in the cluster core using the binary fractions of
  \citet{sollima07} with f$_b^{min} = 0.1$ plotted against the total
  stellar mass contained within the core.  The
  symbols used to indicate the observed and predicted numbers are the
  same as in Figure~\ref{fig:model_best}.  The predicted numbers correspond to
  the best-fitting model parameters, which are f$_{1+1} = 0$,
  f$_{1+2} = 0$, f$_{2+2} = 3.6 \times 10^{-3}$, and f$_{mt} = 1.4
  \times 10^{-3}$.
\label{fig:model_best_rmax_sollima_10min_tBS15}}
\end{figure}

The best fit to the observations is found
by adopting the relation for f$_b$ provided in
Equation~\ref{eqn:sollima07} and setting f$_b^{min} = 0.1$ (however we
note that a comparably good agreement is found with a slightly lower
f$_b^{min} = 0.05$).  This
improves the agreement between our model predictions and the
observations by increasing the number of BSs formed from binary star
evolution in massive clusters.  The result is a good agreement between
our model predictions and the observations in both low- and high-mass
cores, as shown in
Figure~\ref{fig:model_best_rmax_sollima_10min_tBS15}.  In this case,
the best-fitting model parameters yield the lowest deviance of any of
the assumptions so far considered.  
These best-fitting values suggest that most BSs are formed 
from binary star evolution, with a small contribution from 2+2
collisions being needed in order to obtain the best possible match to
the observations.  Similarly to what was found for our initial
assumptions, the ideal contribution from 2+2 collisions
constitutes at most a few percent of the predicted total for most of
the clusters in our sample.  
On the other hand, if we change our imposed minimum binary fraction to
f$_b^{min} = 0.05$ we find that a non-negligible (i.e. up to a few
tens of a percent) contribution from
single-single collisions is needed in several clusters to obtain the best possible
agreement with the observations (which is very nearly as good as was
found using f$_b^{min} = 0.1$).  All of this shows that, although
binary star evolution consistently dominates BS formation in our
best-fitting models, at least some contribution from collisions
(whether it be 1+1 or 2+2 collisions, or some combination of 1+1, 1+2
and 2+2 collisions) also
consistently improves the agreement with the observations.  Moreover,
it is interesting to note that an improved agreement with the
observations could alternatively be obtained if we keep f$_b^{min} =
0.01$ but all or some of
f$_{1+1}$, f$_{1+2}$ and f$_{2+2}$ increase with increasing cluster
mass.  This would also serve to improve the agreement at the high-mass
end.  We will return to this in Section~\ref{discussion}.

Finally, we also tried adopting the observed dependence of f$_b$ on
M$_V$ reported in \citet{milone08}, who also found evidence for an
anti-correlation between the core binary fraction and the total
cluster mass.  Despite this change, we consistently find that our
results are the same as found when using the empirical binary fraction
relation provided in Equation~\ref{eqn:sollima07}.

\subsection{Average BS Lifetime} \label{lifetime}

We also tried changing our assumption for the average BS lifetime.  We
explored a range of plausible lifetimes based on values found
throughout the literature.  Specifically, we explored the range 0.5-5
Gyrs.  We find that at 
the low end of this range, our model fits become increasingly poor.
This is because lower values for $\tau_{BS}$ correspond to smaller
values for r$_{max}$ and decrease the term (1 -
t$_{(dyn),i}$/$\tau_{BS}$) in Equation~\ref{eqn:N-in}.  This reduces the
contribution to the total predicted 
numbers from BSs formed outside the core that fall in via dynamical 
friction.  Conversely, our model fits
improve for $\tau_{BS} > 1.5$ Gyrs since this corresponds to a larger
contribution to N$_{BS}$ from N$_{in}$.  It is important to note,
however, that this same effect can be had by increasing the 
number of BSs formed outside the core, since this would also
serve to increase N$_{in}$ in Equation~\ref{eqn:number-bss}.  This
can be accomplished by, for instance, increasing the binary fraction
outside the core (which would increase the
number of BSs formed from binary star evolution outside the core that
migrate in due to dynamical friction) relative to inside the core.
This seems unlikely, however, given that observations of low-density
globular clusters and open clusters suggest that their binary
fractions tend to drop off rapidly outside the core
\citep[e.g.][]{sollima07}.  We 
will return to this issue in Section~\ref{discussion}.

\begin{figure}
\begin{center}
\includegraphics[width=\columnwidth]{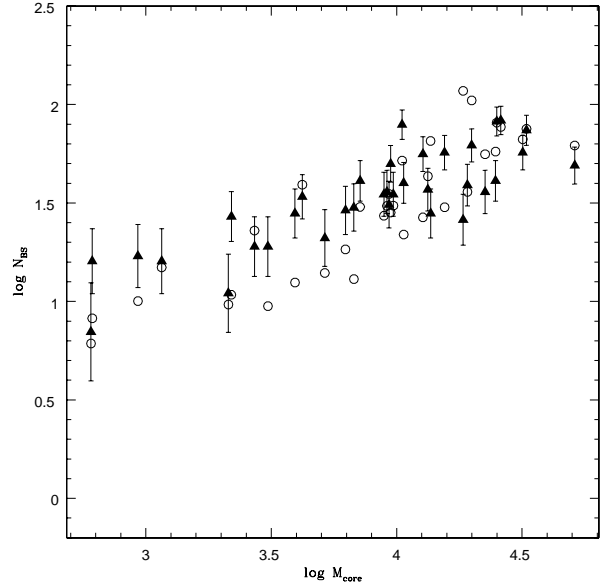}
\end{center}
\caption[The predicted number of BSs plotted versus the total stellar
mass in the core for the best-fitting model parameters found using an
average BS lifetime of 5 Gyrs and the relation for the cluster binary
fraction provided in Equation~\ref{eqn:sollima07} with f$_b^{min} =
0.1$]{The predicted 
  number of BSs plotted versus the total stellar mass in the core for
  the best-fitting model parameters found using an average BS lifetime
  of 5 Gyrs and the relation for the cluster binary fraction provided
  in Equation~\ref{eqn:sollima07} with f$_b^{min} = 0.1$.  The
  symbols used to indicate the observed and predicted numbers are the
  same as in Figure~\ref{fig:model_best}.  The best-fitting model parameters
  used to calculate the predicted numbers are f$_{1+1} = 0$, f$_{1+2}
  = 0$, f$_{2+2} = 1.6 \times 10^{-3}$, and f$_{mt} = 9.9 \times
  10^{-4}$.  The agreement with the observations is excellent for these
  best-fitting values.
\label{fig:model_best_rmax_sollima_10min_tBS50}}
\end{figure}

The best possible match to the observations is
found by adopting an average BS lifetime of 5 Gyrs along with the
relation for f$_b$ provided in Equation~\ref{eqn:sollima07} with 
f$_b^{min} = 0.1$.  The predictions of our model 
are shown in Figure~\ref{fig:model_best_rmax_sollima_10min_tBS50} for
these best-fitting model parameters.  As shown, the agreement
between our model predictions and the observed numbers is excellent.
The best agreement is found by adopting an average BS lifetime of 5
Gyrs, however the agreement is comparably excellent down to 
slightly less than $\tau_{BS} \sim 3$ Gyrs.  Although increasing
$\tau_{BS}$ does contribute to improving the 
agreement between our model predictions and the observations, the
effect is minor compared to the improvement that 
can be found by changing our assumption for the binary fraction.
This is apparent upon comparing 
Figure~\ref{fig:model_best_rmax_sollima_10min_tBS50} to 
Figure~\ref{fig:model_best_rmax_sollima_10min_tBS15}.  

\subsection{Migration} \label{results-migration}

In order to explore the sensitivity of our results to our assumption
for r$_{max}$, we also tried setting N$_{in}$ equal to the total
number of BSs expected to form within 10 r$_c$ for all clusters.  For
comparison, for an average BS lifetime of $\tau_{BS} = 1.5$ Gyrs,
r$_{max}$ ranges from 2 - 15 r$_c$ for the clusters in our
sample.  Despite implementing this change, our results remained the
same.  This is because the largest contribution to the total number of
BSs comes from those BSs formed in the first few shells
immediately outside the core that migrate in due to dynamical
friction.

Several GCs have been reported to show evidence for a decrease in
their binary fractions with increasing distance from the cluster
centre \citep[e.g.][]{sollima07, davis08}.  This effect is often
significant, with binary fractions decreasing by up to a factor of a
few within only a few core radii from the cluster centre.  Based on
this, our assumption that f$_b$ is independent of the distance
from the cluster centre likely results in an over-estimate of the true
binary fraction at large cluster radii.  In order to
quantify the possible implications of this for our results, we tried
setting f$_b$ $= 0$ for all shells outside the core.  Although this
assumption is certainly an under-estimate for the true binary fraction
outside the core, our results remain the same (albeit the agreement
with the observations is considerably worse than for most of our previous
model assumptions).  Once again, the
best-fitting model parameters suggest that most BSs are formed from
binary star evolution, with a non-negligible (i.e. up to a few tens of
a percent in some clusters) contribution from
binary-binary collisions.  Our results indicate that, if the
binary fraction is negligible outside the core, then the contribution
from BSs that migrate into the core due to dynamical friction is also
negligible.  This is because the time between 1+1 collisions increases
rapidly outside the core, and every other BS formation mechanism requires
binary stars to operate.  

We also explored the effects of assuming a non-zero value for
N$_{out}$ in Equation~\ref{eqn:number-bss} by imparting a constant
kick velocity to all
BSs at birth.  Using conservation of energy, we calculated the
cluster radius to which BSs should be kicked upon formation, and used
the time-scale for dynamical friction at the kick radius to calculate
the fraction of BSs expected to migrate back into the core within
a time $\tau_{BS}$.  Regardless of our assumption for the kick
velocity, this did not improve the deviance for any of our
best-fitting model parameters.

\subsection{Average Binary Semi-Major Axis} \label{avg-a}

We investigated the dependence of our results on our assumption for
the average binary semi-major axis.  However, this had a very small
effect on our results.  This is because only N$_{1+2}$ and N$_{2+2}$
depend on the average semi-major axis, and neither of these terms
dominated BS production regardless of our model
assumptions.  Only f$_{2+2}$ is
non-zero for our best-fitting models however, as before, it consistently
suggests that far fewer BSs should be formed from 2+2 collisions than from
binary star evolution. 

\section{Summary \& Discussion} \label{discussion}

In this paper, we have presented an analytic model to investigate BS
formation in globular clusters.  Our model predicts the number of BSs
in the cluster core at the current cluster age by estimating the
number that should have either formed 
there from stellar collisions and binary star evolution, or migrated
in via dynamical friction after forming outside the core.  We have
compared the results of our model predictions for a variety of input
parameters to observed BS numbers in 35 GCs taken from the catalogue of
\citet{leigh11a}.   

What has our model told us about BS formation in dense cluster
environments?  
The agreement between the predictions of our model and the
observations is excellent if we assume that:

\begin{itemize}

\item Binary star evolution dominates BS formation, however at least
  some contribution from 2+2 collisions (most of which
  occur in the core) must also be included in the total predicted
  numbers.  Although it is clear that including a contribution from
  dynamical encounters gives the best possible match to the
  observations, it is not clear how exactly this is accomplished in
  real star clusters.  Does the cluster dynamics increase
  BS numbers via direct collisions?  Or do dynamical interactions
  somehow modify primordial binaries to initiate more mass-transfer
  events?  We will return to this point below.

\item The binary fraction in the core is inversely
  correlated with the total cluster luminosity, similar to the
  empirical relations found by \citet{sollima07} and \citet{milone08}.
  We also require a minimum core binary fraction of $5-10\%$.  The
  inverse dependence of f$_b$ on the total cluster mass
  contributes to a better agreement with the observations at the
  low-mass end of the distribution of cluster masses, whereas 
  the imposed condition that f$_b^{min} = 0.05 - 0.1$ contributes to
  improving the agreement at the high-mass end.

\item BSs formed outside the core that migrate in by the
  current cluster age contribute to the total predicted numbers.

\item The average BS lifetime is roughly a few ($\sim$ 3-5) Gyrs, since this
  increases the fraction of BSs formed outside the core that will have
  sufficient time to migrate in due to dynamical friction.

\end{itemize}

Our model can only provide a
reasonable fit to the observations for all cluster masses if we assume
that the cluster binary fraction is inversely proportional to the total
cluster mass.  It is interesting to consider the possibility that such
an inverse proportionality could 
arise as a result of the fact that the rate of two-body relaxation is
also inversely proportional to the cluster mass \citep{spitzer87}.
Consequently, since binaries tend to be the most massive objects in
GCs, they should quickly migrate into the core in low-mass clusters,
contributing to an increase in the core binary fraction over
time \citep{fregeau09}.  This process should operate on a considerably
longer time-scale in very massive GCs since the time-scale for
two-body relaxation is very long.  Mass segregation could then 
contribute to the observed sub-linear dependence of BS numbers
on the core masses by acting to preferentially migrate the binary star
progenitors of BSs into the cores of low-mass clusters.  This is one
example of how a direct link could arise between the observed
properties of BS 
populations and the dynamical histories of their host clusters.
Although this scenario is interesting to consider, we 
cannot rule out the possiblity that an anti-correlation between the
core binary fraction and the total cluster mass could be a primordial
property characteristic of GCs at birth.

When interpreting our results, it is important to bear in mind that
binary star evolution and dynamical interactions involving binaries
may not always contribute to BS formation independently.  For example,
dynamics could play an important role in changing the distribution of
binary orbital parameters so that mass-transfer occurs more commonly
in some clusters.  
One way to perhaps compensate for this effect would be to include 
a factor of $1/a$ (where $a$ is the average binary semi-major axis) in
Equation~\ref{eqn:N-bin}.  This would serve to account for the fact
that we might naively expect clusters populated by more close binaries
to be 
more likely to have a larger fraction of their binary populations
undergo mass-transfer.  This does not, however, guarantee that more
BSs will form since our poor understanding of binary star evolution
prevents us from being able to predict the outcomes of these
mass-transfer events, and whether or not they will form BSs.
Moreover, little is known about the distribution of orbital
parameters characteristic of the binary populations in globular
clusters, and how they are typically modified by the cluster
dynamics.  For 
these reasons, the interpretation of our results must be done with
care in order to ensure that reliable conclusions can be drawn.

In general, our results suggest that binary stars play a crucial role
in BS formation in dense GCs.  
In order to obtain the best possible agreement with the observations,
an enhancement in BS formation from dynamical encounters is required
in at least some clusters relative to what is expected by assuming a
simple population of binaries evolving in isolation.  It is not clear
from our results, however, how exactly this occurs in real star
clusters.  Dynamics could enhance BS
formation directly by causing stellar collisions, or this could
also occur indirectly if the cluster dynamics somehow induces episodes of
mass-transfer by reducing the orbital 
separations of binaries.  But in which clusters is
BS formation the most strongly influenced by the cluster dynamics?
Unfortunately, no clear trends have emerged from our analysis that
provide a straight-forward answer to this question.  However, our
results are consistent with dynamical interactions 
playing a more significant role in more massive clusters (although
this does not imply that the cluster dynamics does not also contribute
in low-mass clusters).  This could 
be due to the fact that more massive clusters also tend to have higher
central densities \citep[e.g.][]{djorgovski94}, and therefore
higher collision rates.  This picture is, broadly speaking, roughly
consistent with 
the results of \citet{davies04}.  These authors considered the observed
dependence (or lack thereof) of BS numbers on the total cluster masses
presented in \citet{piotto04}, and suggested that primordial binary
evolution and stellar collisions dominate BS production in low- and
high-mass clusters, respectively.

Our model neglects the dynamical evolution of GCs and the resulting
changes to their global properties, including the central density,
velocity dispersion, core radius and binary fraction.  As a young
cluster evolves, dynamical processes like mass segregation and
stellar evaporation tend to result in a smaller, denser core.  Within
a matter of a few half-mass relaxation times, a gravothermal
instability has set in and the collapse ensues on a time-scale
determined by the rate of heat flow out of the core
\citep[e.g.][]{spitzer87}.  We are focussing 
on the last $\tau_{BS}$ years of cluster 
evolution, a sufficiently late period in the lives of most GCs
that gravothermal collapse will have long since taken over as the
primary driving force affecting the stellar concentration in the
core.  Most of the GCs in our sample should currently be in a phase
of core contraction \citep{fregeau09, gieles11}, and their central
densities and core radii should have been steadily
decreasing over the last $\tau_{BS}$ years.  Therefore, by using
the currently observed central cluster parameters and assuming that
they remained constant over the last
$\tau_{BS}$ years, we have effectively calculated upper limits for
the encounter rates.  This could suggest that we have over-estimated
the importance of dynamical interactions for BS formation.  On the
other hand, some theoretical
models of GC evolution suggest that the hard binary fraction in the
core of a dense stellar system will generally increase
with time \citep[e.g.][]{hurley05, fregeau09}.  This can be understood as an
imbalance between the migration of
binaries into the core via mass segregation and the destruction of
binaries in the core via both dynamical encounters and their internal
evolution.  This could suggest that our estimate for the number of BSs
formed from binary star evolution should also be taken as an upper
limit.  The key point is that GC evolution can act to
increase the number of BSs in the cluster core via several different
channels.  The effects we have discussed should typically be small,
however, since 
$\tau_{BS}$ is much shorter than the cluster age \citep{deangeli05}
and any changes to global cluster properties that occur during this
time will often be small.

Our model adopts the same values for all free parameters in all
clusters.  In particular, this is the case for several global cluster
properties, including the average stellar mass, the average BS 
mass, the average BS lifetime, and the average 
binary semi-major axis.  With the exception of the average stellar
mass, there is no conclusive observational or theoretical
evidence to indicate that these parameters should differ from
cluster-to-cluster, although we cannot rule out this possibility.  
For instance, the distribution of binary orbital parameters could
depend on cluster properties 
like the total mass, density or velocity dispersion
\citep[e.g.][]{sigurdsson93}.  In particular, the central velocity
dispersion should be higher in more massive GCs
\citep[e.g.][]{djorgovski94}, which should correspond to a smaller
binary orbital separation for the hard-soft boundary.  
This could contribute to massive GCs tending to have smaller average
binary orbital separations since soft binaries should not survive
for long in the dense cores of GCs \citep[e.g.][]{heggie03}.  In turn,
this could affect the occurrence of mass-transfer events, or of
mergers during 1+2 and 2+2 encounters.  This last point 
follows from the fact that numerical scattering experiments have shown
that the 
probability of mergers occurring during 1+2 and 2+2 interactions
increases with decreasing binary orbital separation
\citep[e.g.][]{fregeau04}.  Both the average stellar mass and the
average BS mass (and hence lifetime) could also depend on the total 
cluster mass, as discussed in \citet{leigh09} and \citet{leigh11a}.

We have also neglected to consider the importance of triples for BS
formation throughout our analysis \citep[e.g.][]{perets09} since we
are unaware of any 
observations of triples in GCs in the literature.  Interestingly,
however, our results for binary star evolution can be generalized to
include the internal evolution of triples since they should have the
same functional dependence on the core mass (i.e. N$_{te} \propto
f_tM_{core}$, where N$_{te}$ is the number of BSs formed from triple
star evolution and f$_{t}$ is the fraction of objects that are
triples). 

Finally, our model 
assumes that several parameters remain constant as a function of
the distance from the cluster centre, including the binary fraction
and the average 
semi-major axis.  However, observations of GCs suggest that their
binary fractions could fall off rapidly outside the core
\citep[e.g.][]{sollima07, davis08}.  Our results suggest that, if the
binary fraction is negligible outside the core, then the contribution
from BSs that migrate into the core due to dynamical friction is also
negligible.  This is because the time between 1+1 collisions increases
rapidly outside the core, and every other BS formation mechanism requires
binary stars to operate.  On the other hand, the presently observed
binary fraction outside the core could be low as a result of binaries
having previously migrated into the core due to dynamical friction
\citep[e.g.][]{fregeau09}.  If these are the binary star progenitors
of the BSs currently populating the core, then dynamical friction
remains an important effect in determining the number of BSs
currently populating the core. 

Despite all of these 
simplifying assumptions, we have shown that our model can reproduce the 
observations with remarkable accuracy.  Notwithstanding, the effects
we have discussed 
could be contributing to cluster-to-cluster differences in the
observed BS numbers.  Our model provides a well-suited resource for
addressing the role played by these effects, however future
observations will be needed in order to obtain the desired constraints
(e.g. binary fractions, distributions of binary orbital parameters, etc.).   

\section*{Acknowledgments}

We would like to thank Ata Sarajedini, Aaron Dotter and Roger Cohen
for providing the observations to which we compared our model
predictions, as well as for providing a great deal of guidance in 
analyzing the data.  We would also like to thank Evert Glebbeek, Bob Mathieu
and Aaron Geller for useful discussions.  This research has been
supported by NSERC and OGS.

\label{lastpage}

\end{document}